\documentclass[preprint,floats,eqsecnum,aps,tightenlines,showpacs]{revtex4}
\usepackage{epsfig}
\usepackage{color}

\begin{document}

\newcommand{\be}{\begin{equation}}
\newcommand{\ee}{\end{equation}}
\newcommand{\bea}{\begin{eqnarray}}
\newcommand{\eea}{\end{eqnarray}}
\newcommand{\nn}{\nonumber}
\newcommand{\dk}{\frac{d^4k}{(2\pi)^4}}
\newcommand{\ep}{i\epsilon}
\newcommand{\dq}{\frac{d^4q}{(2\pi)^4}}
\newcommand{\al}{\alpha}
\newcommand{\om}{\omega}

%

\title{Spectrum of Higgsonium in the SM and beyond }

\author{V.~\v{S}auli}
\affiliation{CFTP and Departamento de F\'{\i}sica,
Instituto Superior T\'ecnico, Av.\ Rovisco Pais, 1049-001 Lisbon,
Portugal, }
\affiliation{Department of Theoretical Physics,
Nuclear Physics Institute, \v{R}e\v{z} near Prague, CZ-25068,
Czech Republic}

\begin{abstract}
Using the formalism of Bethe-Salpeter equation (BSE) the Higgsonium bound state is studied. The condition for formation of Higgsonium bound states are discussed in SM and in the simple extension of.  
\end{abstract}

\maketitle

\section{Introduction}
The Standard Model (SM) of particle physics involves very minimalistic idea of electroweak symmetry breaking scenario. Consequently it results with the only one real scalar field and in fact its last directly unobserved particle- the Higgs boson. However, being inspired of family pattern of SM fermionic sector, it is quite natural to consider the extensions of SM with more rich structure of the scalar sector. By doubling the doublets or/and adding the gauge singlets such next-minimal extensions of SM model have been considered and studied from various perspectives. Clear motivations for such extensions is to reduce some SM shortcomings , like the better agreement with precision electroweak fit, the theoretical problem of mass hierarchy or the dark matter problem.

Having more interacting scalar bosons one can expect  qualitative changes in the scalar boson sector.
In some circumstances the binding forces between scalars can appear strong enough to produce the bound states. What is the spectrum of appearing bound states and how they exhibit their existence in the collider experiments are important questions. Most actually , with the run of LHC, what do we observe if the Higgsonium realizes in the nature?

 As a simplest model  choosen for the actual calculation  is the extension of SM that involves the addition  of a real scalar singlet $S$ to the SM Lagrangian. The phenomenological implications for singlet extension  SM 
(xSM) has been studied from the collider and cosmological perspectives \cite {LHCscalar}. The later typically require small mixing with the SM Higgs and from the perspective of bound states
it reduces to the SM. In such circumstances it was shown in \cite{RUPP} that 
super heavy Higgs $m_H\simeq 1 TeV$ would be needed to form a bound state. In our model we will consider large mixing which could produce two scalars $H_1$ and $H_2$ both having masses at few hundred $GeV$.

\section{Model}

In what follows I will consider the xSM, the SM is  obtained by putting  the new couplings to zero.  
The Lagrange density for the xSM  model is
\begin{equation}
{\cal L}=(D_{\mu}H)^{\dagger}D^{\mu}H+{1 \over 2}\partial_{\mu}S\partial^{\mu}S -V(H,S),
\end{equation}
where $H$ denotes the complex Higgs doublet and $S$ the real scalar. The term linear in $S$ is chosen to vanishes after the spontaneous breaking. The potential is given by
\begin{eqnarray}
&&V(H,S)= {\lambda } (H^{\dagger}H-{v^2 \over 2})^2 +{\delta_1 \over 2}H^{\dagger}H~S  \\
&&+{\delta_2 \over 2}H^{\dagger}H~S^2+ {\delta_1 v^2} S+{\kappa_2 \over 2} S^2 +{\kappa_3 \over 3}S^3+{\kappa_4 \over 4}S^4. \nonumber 
\end{eqnarray}

In unitary gauge the charged component of the Higgs doublet $H$ becomes the longitudinal components of the charged $W$-bosons and the imaginary part of the neutral component becomes the longitudinal component of the $Z$-boson. In Unitary gauge the Higgs field  doublet reads 
\be
\left({\begin{tabular}{c}
   $ 0 $   \\
$\frac{1}{\sqrt{2}}(h+v)$\\   
\end{tabular}}\right)
\ee
Within our notation the mass terms in the scalar potential become
\be
V_{ \rm mass}={1 \over 2}\left(\mu_h^2 h^2 +\mu_S^2 S^2 +\mu_{h S}^2 h S \right),
\ee
where
\begin{eqnarray}
\label{param}
\mu_h^2=2\lambda v^2 \, ,
\mu_S^2= \kappa_2+\delta_2 v^2 \, ,
\mu_{h S}^2=2\delta_1 v.
\end{eqnarray}
The mass eigenstate fields $H_{1,2}$  are linear combinations of the Higgs scalar field $h$ and the singlet scalar field $S$. Explicitly, the inverse transformation reads
\bea
h&=& c ~H_1- s ~H_2\, ;
\nn \\
S&=& s ~H_1+ c ~H_2\, ,
\eea
where $c=cos \theta$, $s=sin \theta$  and the mixing is determined as
 ~\cite{KRASNIKOV}

\begin{equation}
\tan \theta={ x \over 1+\sqrt{1+x^2}},~~~x={\mu_{h S}^2 \over \mu_S^2 - \mu_h^2 }.
\end{equation}
for positive $x$ and for heavier singlet $(x<0) $ we have for mixing angle
\be
\tan \theta={   1+\sqrt{1+x^2} \over |x|}.
\ee

The terms in the scalar potential that break the discrete $S \rightarrow
-S$ symmetry are proportional to the couplings $\delta_1$ and $\kappa_3$ and  
we do not consider these term should very small we assume they are large enough to make a bound state and sufficiently strong communication with the rest of the Standard Model.  
The small mixing scenario with the light singlet like scalar has been considered in 
\cite{KORAWI2006}.  

In this paper we will consider relatively large mixing angle $\theta$ with  Higgsiess masses less then 200 GeV which could be  promising for experimental observation thorough LHC era. For such case the constraints from electroweak precision observables and their implications for LHC Higgs phenomenology has been already analyzed in \cite{LHCscalar}.

The mass eigenstates satisfy in any case
\bea
M_{1}^2&=&\mu_h^2 c^2 +\mu_S^2 s^2+ \mu_{hS}^2 c s  
\nn\\
M_{2}^2&=&\mu_h^2 s^2 +\mu_S^2 c^2- \mu_{hS}^2 c s \, .
\eea

The rest of the scalar potential  contains four new parameters which are added to the SM.
\be
V_{int}=\frac{\lambda}{4}h^4+\frac{\kappa_4}{4}S^4+\lambda v h^3+ \frac{\kappa_3}{3}S^3
+\frac{\delta_2}{2}h^2S^2+\frac{\delta_1}{2}h^2S+\frac{\delta_2}vh S^2
\ee
Based on the nonrelativistic consideration,  the cubic interaction should be sufficiently enhanced against the quartic one, otherwise the bound state cannot be formatted. 

The purely cubic interaction between mass eigenstates can be written in the following way

\be
V_{cub}=g_{111}H_1^3+g_{112}H_1^2H_2+g_{122}H_1H_2^2+g_{222}H_2^3
\ee

with the new couplings $g$`s are  known cubic polynomial of $\sin \theta$
and $cos\theta$.

It is advantageous to evaluate new quartic couplings between physical states :
\be
V_{4}=\frac{g_{111}}{4}H_1^4+\frac{g_{1112}}{4}H_1^3H_2+\frac{g_{1122}}{4}H_1^2H_2^2+\frac{g_{1222}}{4}H_1H_2^3+\frac{g_{2222}}{4}H_2^4
\ee
where $\lambda`$ are known quartic polynomial of cos and sin of mixing angel. 

Higgsonium as any other two body states  in Quantum Field Theory is described by two body BSE:
\be
\Gamma=\int_k VG^{[2]}\Gamma
\ee
where we use shorthand notation $\int_k=i\int\frac{d^4q}{(2\pi)^4}$ and 
$G^{[2]}$ is the two particle propagator of the constituent Higgsies $H_1$.
In momentum space it can be conventionally written as
\bea
G^{[2]}(k,P)&=&D(k+P/2,M_1^2)D(-k+P/2,M_1^2) \,  ;
 \\
D(k,M^2)&=&\frac{1}{k^2-M^2-\ep} \, ,
\eea

Let assume that the attractive interaction between heavy Higgsiess $H_2$ is
strong enough to form a bound state. Within the xSM  the irreducible BSE kernel in lowest order reads
\be \label{tree}
V=6\lambda_{1111}+\sum_{x=s,t,u}\left[\frac{4 g_{112}^2}{x-M_2^2}+\frac{36 g_{111}^2}{x-M_1^2}\right] \,
\ee
where the first term represent the pure constant interaction and the $s,t,u$ are usual Mandelstam variables. Down indices show which field - $H_1$ or/and $H_2$ belong to a given interaction vertex.

It is advantageous to explicitly divide the solution which is independent on the relative momenta
of the constituents. Following the notation in \cite{RUPP} the original BSE can be rewritten in the following form
\be \label{BSE}
\Gamma_p(p,P)=\Gamma_I(P)\int_k V_p(k,p,P) G^{[2]}(k,P)+\int_k V_p(k,p,P) G^{[2]}(k,P)\Gamma_p(k,P)\, ,
\ee
where  
\bea
 V_I=V_c+V_s \, , \, V_p&=&V_t+V_u \, ,
\eea
The first term is supposed to collect all constant term. i.e the ones the do not depend on the relative momenta.  In our tree kernel approximation reads
\be \label{const}
V_I=6\lambda_{1111}+\frac{4 g_{221}^2}{P^2-M_1^2}+\frac{36 g_{222}^2}{P^2-M_2^2} \,
\ee
where the full solution of BSE is given by the sum
\be
\Gamma(p,P)=\Gamma_I(P)+\Gamma_p(p,P)
\ee
The equation for  the function $\Gamma_I(P)$ is purely algebraic
\be  \label{druha}
\Gamma_I(P)=\frac{V_I \int_k  \Gamma_p(k,P)G^{[2]}(k,P)}{1-V_I\int_k G^{[2]}(k,P)}
\ee

The BSE represents the singular equation which can be solved by some known method.
The one known possibility is to perform a Wick rotation for a relative momenta of constituents while keeping the total square of four momenta $P^2$ timelike.

The other well known possibility is the Minkowski solution performed within the utilization  of the unique integral representation  of the kernels and amplitudes that appear in the BSE \cite{NAKAN,SAULIphd}.

The bound state vertex function can be expressed as the following   

\bea \label{bounstate}
\Gamma(P,p)=\int\limits_{-1}^1 d\eta\,
\int\limits_{-\infty}^{\infty}d\alpha\,
\frac{\rho^{[n]}(\alpha,\eta)}{[F(\al,\eta;P,p)]^n} \, ,
\eea

where $n$ is a integer and all the singularity is trapped by the zeros of the denominator in (\ref{bounstate}) which reads 

\be
F(\al,\eta;P,p)=\al-(p^2+P.p z+\frac{P^2}{4})-\ep
\ee

Recall the known  property of superrenormalized model studied yet: The function $ \rho^{[n]}$  is more smooth as larger $n$ is used. In practise, the BSE were solved for generalized Wick-Cutkosky models for the lowest value only, i.e.   $n=1,2$. The studied models were 
very simple  ones with the most cubic scalar interaction presented.

 Here, having the quartic interaction presented, the inhomogeneous term is generated represented by $\Gamma_I(P)$ which is just real constant for a given discrete value of bound state mass $P$.
To avoid more complicated distribution we naturally assume

\bea \label{bound2}
\Gamma(P,p)=\Gamma_I(P)+\int\limits_{-1}^1 d\eta\,
\int\limits_{-\infty}^{\infty}d\alpha\,
\frac{\rho_p(\alpha,\eta)}{[F(\al,\eta;P,p)]} \, ,
\eea
where $\rho_p(\alpha,\eta)$ is assumed to be a real function, and not delta distribution. It fully corresponds to the function $\Gamma_p$, noting its structure is fully driven by pure triplet interaction of Higgs. Furthermore, we explicitly choose $n=1$ in the IR  (\ref{bound2}), following 
the most easy integral representation of the inhomogeneous term 
in the expression, i.e.  $\int_k V_p(k,p,P) G^{[2]}(k,P)$. These integral corresponds to the Feynman scalar triangle diagram. 

One can show that the BSE can be converted to the regular integral equation for $\rho_p$. It reads
\be \label{vysledek}
\rho_p(\alpha,\eta)= \frac{1}{\al-M_1^2}\left[\Gamma_I(P)\rho_{I}(\alpha,\eta)+
 \int\limits_{-1}^1 d z\, \int\limits_{-\infty}^{\infty}d a \,\rho_p(a,z) 
{\cal V}(\alpha,\eta,a,z)\right]
\ee
where $\rho_I, {\cal V}$ are  known regular function and the   
constant term arises due to the quartic interaction

\be
\Gamma_I(P)=\frac{V_I^R  \int\limits_{-1}^1 d z\, \int\limits_{-\infty}^{\infty}d a
\rho_p(a,z) I_{F}(P^2;a,z)}{1-V_I^R I_B^{[R]}(P^2)}
\ee
where $V_I^R$ is the renormalized constant interaction and  $I_F$,  $I_B$ stand for the known one loop triangle and one loop bubble integrals.

To renormalize the momentum subtraction renormalization scheme with zero momentum scale is used, 
\be
V_I^R=6\lambda_{1111}^R+\frac{4g_{112}^2}{P^2-M_2^2}+\frac{36g_{111}^2}{P^2-M_1^2}
\ee
where $\lambda_{1111}^R$ is the renormalized quartic coupling of heavier Higgs mass eigenstates
$H_1$.

\section{Results}

After a suitable normalization then the BSE for the weight function (\ref{vysledek}) has been solved by the method of iteration. The coupling constants have been varied to reach the real discrete spectrum of Higgsonia.

Firstly I mention the result for SM Higgs. The only known input is the Higgs vev. $v=275 GeV$ while the Higgs mass and the cubic coupling  depends on the experimentally unknown $\lambda$, these satisfy $m_h=\sqrt{2\lambda}v, \lambda_3=2\lambda v$. There are no bound states bellow certain critical coupling of $\lambda_3$. The first Higgsonium of the mass $M=2m_h$ is formed when  $m_h=1.3 TeV$.   Such a result doesn't provide reliable answer, moreover it brings more questions. 

Such a heavy Higgs boson is ruled out by the electroweak precision test. Furthermore, for such a fat Higgs, the Higgs sector of SM represents strongly coupled field theory  and our BSE becomes only rough estimate of reality. In addition, switching on the top quark Yukawa, the fat Higgs becomes broad resonance and its fast decays should  prevent the formation for bound states.

To have a reasonable model which is not completely  ruled out by electroweak oblique correction constraints we assumed relatively light scalars in the xSM. There is no bound state unless the new cubic coupling is  large enough. To compare various quantities, the all dimensionfull quantities are scaled in units of Higgs vev. $v$.  Then roughly pronounced, the new cubic coupling must be several times larger then the rest of Lagrangian parameters. This is the  main conclusion from the numerical inspection of the large regime of xSM parameter space.  

Here I present the first preliminary numerical result: The parameters used as an input were as the following:
 $v=275 GeV, \, \lambda=0.20, \, \delta_1=1.20*v, \, \delta_2=0.40$,
$\kappa_2=0.10*v, \,   \kappa_3=5.0*v , \, \kappa_4=0.20$.
 
It leads to two massive eigenstates with $M_1=179.5 GeV, \, M_2=177.7 GeV,$ and couplings $g_{111}\simeq 280 GeV,g_{222}\simeq 400 GeV $ , 
$\lambda_{1111}\simeq 0.33$ and the appropriate mixing 
$cos\theta=0.696$. Solution of BSE gives the bound states 20 $\%$ lighter than the Higgs production threshold :
\be
M_B=0.8*2*M_1=286 GeV\, .
\ee

\end{document}